\documentstyle[twocolumn,pra,aps,epsfig]{revtex}
\begin{document}
\flushbottom
\draft
\title{Eliminating the mean-field shift in multicomponent Bose-Einstein
condensates}
\author{E. V. Goldstein, M. G. Moore, H. Pu and P. Meystre}
\address{Optical Sciences Center,
University of Arizona, Tucson, Arizona 85721\\ (May 1, 2000)
\\ \medskip}\author{\small\parbox{14.2cm}{\small\hspace*{3mm}
We demonstrate that the nonlinear mean-field shift in a
multi-component Bose-Einstein condensate may be eliminated by
controlling the two-body interaction coefficients.
This modification is achieved by, e.g., suitably engineering
the environment of the condensate.
We consider as an example the case
of a two-component condensate in a tightly confining atom
waveguide. Modification of the atom-atom interactions is then
achieved by varying independently the transverse wave function of
the two components. Eliminating the density dependent phase shift
in a high-density atomic beam has important applications in atom
interferometry and precision measurement.\\
\\[3pt]PACS numbers: 03.75.Fi, 03.75.Be, 03.75.-b }}
\maketitle \narrowtext The promise of Bose-Einstein condensates
\cite{And95,Dav95} and atom lasers
\cite{Mew97,Hag99,Blo99,AndKas99}as sources for atom
interferometry \cite{Ber96} based sensors\cite{GusBouKas97}
results from their high brightness and coherence which leads to an
increase in the signal-to-noise ratio as compared to conventional
atom optics. Additionally, further enhancement of sensitivity
might be achieved by taking advantage of nonlinear effects which
occur in quantum-degenerate atomic fields. For example, feedback
between optical and/or matter-wave fields can result in nonlinear
instabilities\cite{MooMey99a,MooMey99b}. Such positive feedback
between optical and atomic intensity gratings has already led to
the design of matter-wave
amplifiers\cite{LawBig98,MooMey99a,Ino99,Koz99}. The utility of
such instabilities for nonlinear interferometry lies in the
associated increase in sensitivity that can be achieved by
operating just above or just below the instability threshold. This
allows a small perturbation to produce a large change in the
properties of the system.

The transition to high density atomic samples carries a price,
however, in that atom-atom collisions may introduce unwanted
nonlinear (density dependent) phase-front distortions which limit
sensitivity. While collisions are often viewed as a source of
decoherence in quantum optics, in the regime of coherent atomic
matter waves atom-atom interactions lead to coherent nonlinear
wave-mixing and can therefore be manipulated using techniques
inspired from nonlinear optics. In this Letter we discuss how the
density-dependent phase shift can be eliminated in a trapped
multi-component condensate. With applications in atom
interferometry in mind we consider specifically the case of a
narrow atomic waveguide such as those recently microfabricated on
glass chips \cite{MulAndGro99,DekLeeLor00}, as these devices hold
great promise for the development of integrated
atom-interferometric devices.

At temperature $T\simeq0$ a single-component Bose-Einstein
condensate is characterized by a scalar order parameter whose
evolution is governed by a nonlinear Schr\"odinger equation
(NLSE). At densities low enough that three-body collisions can be
neglected the NLSE contains a cubic nonlinearity whose form is
determined by the two-body collision potential. In the limit of
$s$-wave scattering, the potential is of the form $V = (4\pi
\hbar^2 na/m) \delta({\bf r}_{12})$, where $m$ is the particle
mass, $n$ the atomic number density, ${\bf r}_{12}$ the relative
position of the atoms, and $a$ the $s$-wave scattering length.
This leads to a density-dependent phase shift in the evolution of
the condensate wave function. In the language of nonlinear optics
this process is known as {\em self-phase modulation}. Since the
exact density of the condensate is usually not perfectly known
(especially in cases where it has been divided by a
`beam-splitter') this is a source of uncertainty that limits the
accuracy with which precision measurements can be done.

There is at first sight nothing that can be done to eliminate this
shift short of reducing the condensate density to a point where it
becomes negligible, or of taking advantage of Feshbach resonances,
in which case three-body collisions appear to become a serious
problem\cite{feshbach}. The situation is quite different, however,
for {\em multi-component} condensates. In addition to a self-phase
modulation proportional to its own density, each condensate
component experiences in that case an additional {\em cross-phase
modulation}, i.e. a phase shift proportional to the density of the
other component\cite{ho,esry,elena,PuBig98a,PuBig98b}. As we will
demonstrate, it is possible to engineer the environment of the BEC
so that the phase shifts associated with self- and cross-phase
modulation cancel each other, thus eliminating the
density-dependent mean-field shift from the condensate evolution.
As the property which governs the evolution of the condensate
phase is the chemical potential, the elimination of the mean-field
shift is equivalent to having a chemical potential which is
independent of the number of atoms in the BEC. In addition, we
show that it is possible for at least one branch of the
quasiparticle spectrum to be density-independent as well.
Measurements which excite only the `collisionless' branch are
therefore insensitive to nonlinear phase shifts and hence of
considerable interest in atom interferometry.

To illustrate how this works we consider the simplest possible situation
that can lead to the required effect, namely
a two-component condensate described by the many-body Hamiltonian
\begin{eqnarray}
    \hat{H} &=& \sum_{j=a,b} \int d^3r \,{\hat \psi}_j^\dag({\bf r})
    \left [-\frac{\hbar^2}{2m} \nabla^2 +U_j({\bf r})\right]
    {\hat\psi}_j({\bf r}) \nonumber \\
    &+&\frac12\sum_{j=a,b} \hbar g_j \int d^3r\, {\hat \psi}_j^\dag({\bf r})
    {\hat \psi}_j^\dag({\bf r}) {\hat \psi}_j({\bf r}){\hat \psi}_j({\bf r})
\nonumber \\
    &+& \hbar g_x \int d^3r\, {\hat \psi}_a^\dag({\bf r})
    {\hat \psi}_b^\dag({\bf r}) {\hat \psi}_b({\bf r}){\hat \psi}_a({\bf r}).
\label{h}
\end{eqnarray}
Here, $U_j$ includes both the external trapping potential
and the internal atomic energy for component $j$,
and the constants $g_a$, $g_b$, and $g_x$ give the strengths of
the nonlinearities due to atom-atom collisions. In the language of
nonlinear optics, $g_a$ and $g_b$ determine the self-phase
modulation of the two components, whereas $g_x$ governs
cross-phase modulation between them.

The specific physical system that we have in mind consists of a
condensate in a one-dimensional atomic waveguide with tight
transverse confinement in the $x$-$y$ plane, i.e., $U_j({\bf
r})=U_j(x,y)$, while propagation along the $z$-dimension is free.
The two components are two internal states (e.g., the Zeeman
sublevels) of the same atomic species. We assume that they may
convert into each other through a weak linear coupling. The ground
state wave functions $\psi_j({\bf r})$ are hence found by
minimizing the Hartree energy functional while holding the total
number of atoms fixed (although the individual atom numbers may
vary). This linear coupling is either sufficiently weak, or is
turned off adiabatically after the steady state has been
established, that we do not explicitly include it in the
Hamiltonian.

As ideally one would like to have a `single-mode' wave guide, we
assume that the transverse confinement is sufficiently strong that
all atoms are `frozen' in the ground state $\varphi_j(x,y)$ of the
transverse potential $U_j(x,y)$. This leads us to introduce the
atomic annihilation operators $ \hat{\phi}_j(z,t)$ for atoms in
the transverse ground state as
\begin{equation}
    \hat{\phi}_j(z,t)=\int
    dxdy\,\varphi_j^\ast(x,y)\hat{\psi}_j({\bf r},t).
\label{defphi}
\end{equation}
From the Hamiltonian (\ref{h}) the Heisenberg equation of motion
for $\hat{\phi}_j(z)$ is found to be
\begin{eqnarray}
    \frac{d}{dt}\hat{\phi}_j(z)&=&
    -i\left[-\frac{\hbar}{2m}\frac{\partial^2}{\partial z^2}+E_j\right]
    \hat{\phi}_j(z)\nonumber\\
    &-&i\left[V_j\hat{\phi}^\dag_j(z)\hat{\phi}_j(z)
    +V_x\hat{\phi}^\dag_k(z)\hat{\phi}_k(z)\right]\hat{\phi}_j(z),
\label{dphijdt}
\end{eqnarray}
where $k\neq j$ and
\begin{eqnarray}
    V_j &=& g_j\,\int\,dxdy\,|\varphi_j(x,y)|^4\,;\;\;\;j=a,b \\
    V_x &=& g_x\,\int\,dxdy\, |\varphi_a(x,y)|^2\,|\varphi_b(x,y)|^2.
\label{defV}
\end{eqnarray}
We choose the energy reference such that $E_a=-\delta/2$ and
$E_b=\delta/2$. From these expressions it is immediately apparent
that the effective phase modulation constants along the
$z$-dimension can be modified by appropriately varying the
transverse trapping potential $U_j(x,y)$.

Assuming that any perturbation present is too weak to excite the
transverse degrees of freedom, the ground state wave function and
quasiparticle spectrum of low-lying excitations are found by
decomposing the boson field operator $\hat{\phi}_j(z,t)$ around
the ground state Hartree wave function as
\begin{equation}
\hat{\phi}_j(z,t)=\left[ \phi_j(z)+\delta\hat{\phi}_j(z,t) \right]
\,e^{-i\omega_0t}, \label{linear}
\end{equation}
where the perturbation operators $\delta\hat{\phi}_j(z,t)$ satisfy
the boson commutation relations:
\begin{equation}
\left[
\delta\hat{\phi}_j(z,t),\,\delta\hat{\phi}_k^{\dagger}(z',t)
\right]=\delta_{jk}\delta(z-z').
\end{equation}
Substituting (\ref{linear}) into (\ref{dphijdt}) and keeping only
the leading-order terms leads to a time-independent
Gross-Pitaevskii equation from which the ground-state wave
function $\phi_j(z)$ and chemical potential $\hbar\omega_0$ can be
determined.

We work in the condensate rest frame and seek plane wave solutions
corresponding to a uniform beam of atoms moving along the wave
guide. The densities of the two components are then given by
\begin{eqnarray}
\rho_a\equiv|\phi_a|^2&=&\frac{(V_b-V_x)\rho -\delta}{V_a+V_b-2V_x},
\nonumber
\\ \rho_b\equiv|\phi_b|^2&=&\frac{(V_a-V_x)\rho +\delta}{V_a+V_b-2V_x},
\label{popul}
\end{eqnarray}
where we have assumed a fixed total density $\rho=\rho_a+\rho_b$.
The relative phase between the two components is arbitrary, as it
is either random, or fixed by the linear coupling whose strength
we have assumed to be negligible. The frequency of phase rotation
for the ground state (chemical potential divided by $\hbar$) is
determined to be
\begin{equation}
\omega_0=\frac{2(V_aV_b-V_x^2)\rho+(V_b-V_a)\delta}{2(V_a+V_b-2V_x)}.
\label{om0}
\end{equation}
In the case $V_a=V_b$, Eqs.~(\ref{popul}) and (\ref{om0})
reduce to Eqs.~(8) and (12) of Ref.~\cite{elena}, respectively.

The phase-rotation frequency $\omega_0$ of the condensate ground
state contains two contributions, one being proportional to the
total density $\rho$, and the other to the detuning $\delta$. Both
terms result from the combined effects of cross- and
self-modulation. In contrast to the case of a single-component
condensate, Eq. (\ref{om0}) shows that  the cross and self-phase
modulation contributions can conspire to cancel the
density-dependent term, provided only that the condition
\begin{equation}
 V_x^2 = V_aV_b\;\;\;{\rm and}\;\;\;V_a+V_b-2V_x\neq 0
\label{cond1}
\end{equation}
is met.
When the solutions given by Eq. (\ref{popul})
give negative densities no homogeneous ground state exists.
Thus we must verify whether or not these conditions
can be fulfilled for positive densities $\rho_a$ and $\rho_b$.
We consider the most common situation where all
two-body interactions are repulsive, $V_a, V_b, V_x>0$, and assume without
loss of generality that $V_b> V_x$, which implies from Eq.~(\ref{cond1})
that $V_a < V_x$. It is then easily shown that the requirement of
$\rho_a >0$ and
$\rho_b > 0$ yields
\begin{equation}
\frac{V_x}{V_b} < \frac{\delta}{(V_b-V_x)\rho} < 1,
\end{equation}
which can be achieved by an appropriate choice of the detuning $\delta$ and/or
of the condensate density $\rho$.

To show that these conditions can be met in a real system by
modifying the trapping potential, we consider the two-component
$^{87}$Rb condensate, where the components $a$ and $b$ correspond
to the hyperfine Zeeman sublevels $|F=2,m_f=1\rangle$ and
$|F=1,m_f=-1\rangle$, respectively. The phase modulation constants
are in the ratio $g_a:g_x:g_b=0.97:1.00:1.03$. We consider the
case of harmonic trapping potentials $U_a=m\omega^2(x^2+y^2)/2$
and $U_b=m\omega^2[(x-x_0)^2+y^2]/2$, the offset $x_0$ in their
centers $x_0$ being a control variable that can be changed via a
bias magnetic field. We have then
\begin{eqnarray}
V_{a,b} &=& \frac{1}{2\pi\xi^2}\,g_{a,b} \nonumber \\
V_x &=& \frac{1}{2\pi\xi^2} \,e^{-x_0^2/2\xi^2}\,g_x
\end{eqnarray}
where $\xi=\sqrt{\hbar/m\omega}$ is the extension of the trap
ground state. Hence to have condition (\ref{cond1})
satisfied, one can choose $x_0=0.03\xi$.

When the condition (\ref{cond1}) is satisfied,
we obtain the remarkable result that the evolution of the
condensate phase is independent of density. We emphasize that by
itself, this is of course not sufficient to guarantee measurements
unperturbed by nonlinear phase shifts can be carried out. Indeed,
the detection of any signal relies on some departure of the state
of the condensate away from its ground state. Assuming that this
change is small, we can describe it in a linearized approach
following the Bogoliubov treatment. We proceed by expanding the
Hamiltonian (\ref{h}) with the help of (\ref{defphi}) and
(\ref{linear}), keeping only quadratic terms in the operators
$\delta\hat{\phi}_j$, yielding
\begin{eqnarray}
    \hat{H}&\approx&\sum_{j=a,b}\int dz\,
    \delta\hat{\phi}^\dag_j(z)
\left[-\frac{\hbar^2}{2m}\frac{\partial^2}{\partial z^2}
    +\hbar V_j|\phi_j|^2 \right]\delta\hat{\phi}_j(z)\nonumber\\
    &+&\sum_{j=a,b}\frac{\hbar}{2}V_j\int dz
    \left[\phi^2_j\delta\hat{\phi}^\dag_j(z)\delta\hat{\phi}^\dag_j(z)
+H.c.\right]\nonumber\\
    &+&\hbar V_x\int dz\Big[\phi^\ast_a\phi^\ast_b\delta\hat{\phi}_a(z)
    \delta\hat{\phi}_b(z) +\phi_a\phi^\ast_b\delta\hat{\phi}^\dag_a(z)
    \delta\hat{\phi}_b(z) \nonumber\\ &&\quad\quad\quad +H.c.\Big].
\label{hquad}
\end{eqnarray}
No first-order contributions in $\delta\hat{\phi}(z)$ appear in
Eq.~(\ref{hquad}) as a consequence of the fact that the condensate
spinor $\phi_j(z)$ satisfies the time-independent nonlinear
Schr\"odinger equation.

The effective Hamiltonian Eq.(\ref{hquad}) can be diagonalized via
a generalized Bogoliubov transformation. As there is no confining
potential in the $z$-direction, we expand the operators
$\delta\hat{\phi}_j(z)$ onto plane waves as
\begin{equation}
    \delta\hat{\phi}_j(z)=(2\pi)^{-1/2}\int dk \,e^{ikz}\, \hat{c}_j(k).
\end{equation}
As a consequence of momentum conservation, only the operators
$\hat{c}_j(k)$, $\hat{c}^\dag_j(k)$, $\hat{c}_j(-k)$, and
$\hat{c}^\dag_j(-k)$ are coupled. In order to diagonalize the
Hamiltonian (\ref{hquad}) we introduce the annihilation operators
for quasiparticles with well-defined momentum $k$ according to
\begin{equation}
    \hat{b}_\mu(k)=\sum_{j}\left[u_{\mu j}(k)\hat{c}_{j}(k)
    +v_{\mu j}(k)\hat{c}^\dag_j(-k)\right]. \label{bjk}
\end{equation}
Invariance under rotation of the coordinate axes clearly requires
that $u_{\mu j}(-k)=u_{\mu j}(k)$ and $v_{\mu j}(-k)=v_{\mu
j}(k)$. The coefficients $u_{\mu j}(k)$ and $v_{\mu j}(k)$ are
determined by the requirements that the operators $\hat{b}_\mu(k)$
and $\hat{b}^\dag_\mu(k)$ obey the bosonic commutation relations
\begin{equation}
    \left[\hat{b}_\mu(k),\hat{b}_\nu(k')\right]=0
\label{bcomm1}
\end{equation}
and
\begin{equation}
    \left[\hat{b}_\mu(k),\hat{b}_\nu^{\dagger}(k')\right]=
    \delta_{\mu\nu}\delta(k-k'),
\label{bcomm2}
\end{equation}
and that the Hamiltonian (\ref{hquad}) takes the form
\begin{equation}
    \hat{H}=\sum_\mu\int dk \,\hbar\omega_\mu({k})\hat{b}^\dag_\mu(k)
    \hat{b}_\mu(k),
\label{bHam}
\end{equation}
where the $\omega_\mu(k)$ are thus the frequencies of the
elementary modes for small collective excitations of the
condensate.

For a fixed momentum $k$ the coefficients $u_{\mu j}(k)$ and
$v_{\mu j}(k)$ may be represented as the matrix elements of the
$2\times 2$ matrices ${\bf U}$ and ${\bf V}$. In order to satisfy
the commutation relations (\ref{bcomm1}) and (\ref{bcomm2}) it is
sufficient that
\begin{equation}
    {\bf UU}^\dagger-{\bf VV}^\dagger={\bf I}, \mbox {\hspace{1.5cm}} {\bf
UV}^T=
    {\bf VU}^T.
\label{uv}
\end{equation}
The functions $u_{\mu j}(k)$ and $v_{\mu j}(k)$ can then be
determined by substituting Eq.(\ref{bjk}) into the commutator
$[b_\mu(k),H]=\hbar\omega_\mu(k) b_\mu(k)$, which guarantees that
the Hamiltonian is of the form (\ref{bHam}). This yields:
\begin{equation}
\omega_{\mu}\sigma_{\mu}={\bf M} \,\sigma_{\mu}
\label{excitation}
\end{equation}
where $\sigma_{\mu} \equiv (u_{\mu a},\,u_{\mu b},\,-v_{\mu a},\,
-v_{\mu b})^T$ and the matrix ${\bf M}$ is given by
\[
{\bf M}=\left( \begin{array}{cccc}
H_a & V_x\phi^*_a\phi_b & V_a(\phi^*_a)^2 &
V_x \phi_a^*\phi_b^* \\
V_x \phi_a \phi_b^* & H_b & V_x\phi_a^*\phi_b^* &
V_b(\phi_b^*)^2 \\
-V_a\phi_a^2 & -V_x \phi_a\phi_b & -H_a &
-V_x\phi_a\phi_b^* \\
-V_x \phi_a\phi_b & -V_b \phi_b^2 & -V_x \phi_a^*\phi_b &
-H_b \end{array} \right)
\]
where $H_j=\hbar k^2/(2m)+V_j \rho_j$.
The eigenfrequencies
$\omega_{\mu}(k)$ whose corresponding eigenvectors satisfy
Eqs.~(\ref{uv}) are
\begin{eqnarray}
\omega_\mu(k) &=&\left\{\frac{\hbar k^2}{2m}\left[
\left(\frac{\hbar k^2}{2m}+V_a\rho_a + V_b\rho_b\right) \right .
\right . \nonumber
\\ &\pm& \left . \left . \left[ (V_a\rho_a-V_b\rho_b)^2+4V_{x}^2
\rho_a\rho_b\right]^{1/2}\right]\right\}^{1/2} . \label{spectrumR}
\end{eqnarray}
These solutions therefore
yield the two branches of the quasiparticle excitation spectrum
for the two-component condensate.

It is straightforward to show that one of the branches does not
depend on the density $\rho$ when the condition (\ref{cond1}) is
satisfied. In
this case the eigenfrequencies are \begin{eqnarray}
    \omega_+(k) & = & \sqrt{\frac{\hbar k^2}{2m}\left [ \frac{\hbar
    k^2}{2m}+2(V_a\rho_a+V_b\rho_b)\right ]} \nonumber \\
    \omega_-(k)& = & \frac{\hbar k^2}{2m},
    \label{spect1}
\end{eqnarray}
hence $\omega_-$ corresponds to the `collisionless' branch.

The existence of a collisionless branch is related to the
invariance of the system under translation along the $z$-axis. It is
easy to show that the corresponding eigenvector in
this branch is given by
\[ \sigma_-(k)=(V_x \phi_b,\,-V_a \phi_a,\,0,\,0)^T \]
which is $k$-independent. The condensate wave function is a plane
wave, but the choice of inertial frame in which this plane wave is
at rest $(k=0)$ is arbitrary. The cancellation of self- and
cross-phase modulation is achieved by having a uniform density
along the $z$-axis, with a constant ratio of the components $a$
and $b$. If a fraction of the total atoms are boosted into a
moving frame while still maintaining their internal superposition
state, then the relative densities of the two components are not
changed. This leads to the possibility of imparting kinetic energy
onto the system without affecting the balance required to
eliminate density-dependent phase shifts. As the coefficients
$v_{-j}$ are found to be zero, it is clear that the quasiparticles
created by $b_-^\dag(k)$ correspond simply to condensate atoms
boosted into a different momentum eigenstate. This remarkable
property, reflected in the collisionless branch of (\ref{spect1})
should be particularly useful for atom interferometry, where the
ability to use Bragg-pulses\cite{bragg1,bragg2} to induce
transitions between the condensate state and the collisionless
branch could lead to the design of beam-splitters which maintain
mean-field-free conditions.

In conclusion, we have shown that for certain values of
collisional constants, the nonlinear phase shift in a
two-component Bose condensate can be completely eliminated. Such
an effect results from the interplay between self- and cross-phase
modulation between the two components. While we have explicitly
shown how it is possible to adjust the values of these collisional
constants in a one-dimensional atomic waveguide, the same method
also applies to two-dimensional systems with strong confinement in
the third dimension. We expect that these results will play an
important role in atom interferometry where uncontrollable
nonlinear phase shift limits applications in precision
measurement.

Finally, the question remains to determine whether the elimination
of the mean-field shifts can be achieved in a more typical 3-d
trap as well. In that case, the ground state must be
determined from a full three-dimensional time-independent set of
Gross-Pitaevskii equations. It will be interesting to see if
the chemical
potential can be made density-independent in this case by appropriately
engineering the
trapping potentials of the different components. Because the existence of the
collisionless branch of the excitation spectrum appears to be related to
translational invariance, future studies will be required to determine whether
such a branch exists in the case of a 3-dimensional trap as well.

This work is supported in part by Office of Naval Research
Research Contract No. 14-91-J1205, National Science Foundation
Grant PHY98-01099, the Army Research Office and the Joint Services
Optics Program.


\begin{references}
\bibitem{And95}M. H. Anderson {\em et al.}, Science {\bf 269}, 198 (1995)
\bibitem{Dav95}K. B. Davis {\em et al.}, Phys. Rev. Lett. {\bf 75}, 3969 (1995).
\bibitem{Mew97}M.-O. Mewes {\em et al.}, Phys. Rev. Lett. {\bf 78}, 582 (1997).
\bibitem{Hag99}E. W. Hagley {\em et al.}, Science {\bf 283}, 1706 (1999)
\bibitem{Blo99}I. Bloch, T. W. H\"ansch, and T. Esslinger, Phys. Rev. Lett. {\bf 82}, 3008 (1999).
\bibitem{AndKas99} B. Anderson and M. A. Kasevich, Science {\bf 282}, 1686 (1999).
\bibitem{Ber96} P. R. Berman (ed.), {\em Atom Interferometry} (Academic, Boston, 1996).
\bibitem{GusBouKas97} T. L. Gustavson, P. Bouyer, and M. A. Kasevich, Phys. Rev. Lett. {\bf 78},
2046 (1997)
\bibitem{MooMey99a} M. G. Moore and P. Meystre, Phys. Rev. A {\bf 59}, R1754 (1999).
\bibitem{MooMey99b} M. G. Moore and P. Meystre, Phys. Rev. A {\bf 59}, R1754 (1999).
\bibitem{LawBig98} C. K. Law and N. P. Bigelow, Phys. Rev. A {\bf 58}, 4791 (1998).
\bibitem{Ino99}S. Inouye {\em et al.}, Science {\bf 285}, 571 (1999).
\bibitem{Koz99}M. Kozuma {\em et al.}, Science {\bf 286}, 5448 (1999).
\bibitem{MulAndGro99} D. Muller {\em et al.}, Phys. Rev. Lett. {\bf 83}, 5194 (1999).
\bibitem{DekLeeLor00} N. H. Dekker {\em et al.}, Phys. Rev. Lett. {\bf 84}, 1124 (2000).
\bibitem{feshbach}J. Stenger {\em et al.}, Phys. Rev. Lett.
{\bf 82}, 2422 (1999).
\bibitem{ho}Tin-Lun Ho and V. B. Shenoy, Phys. Rev. Lett.
{\bf 77}, 3276 (1996).
\bibitem{esry}B. D. Esry {\em et al.}, Phys. Rev. Lett.
{\bf 78}, 3594 (1997).
\bibitem{elena}Elena V. Goldstein and P. Meystre, Phys. Rev. A
{\bf 55}, 2935 (1997).
\bibitem{PuBig98a}H. Pu and N. P. Bigelow, Phys. Rev. Lett.
{\bf 80}, 1130 (1998).
\bibitem{PuBig98b}H. Pu and N. P. Bigelow, Phys. Rev. Lett. {\bf 80}, 1134 (1998).
\bibitem{bragg1}M. Kozuma {\em et al.}, Phys. Rev. Lett.
{\bf 82}, 871 (1999).
\bibitem{bragg2}J. Stenger {\em et al.}, Phys. Rev. Lett.
{\bf 82}, 4569 (1999).
\end{references}
\end{document}